\documentclass[10pt, aps,pra,showpacs,twocolumn, longbibliography]{revtex4-2}

\usepackage[dvipdfmx]{graphicx}
\usepackage{dcolumn}
\usepackage{bm}
\usepackage{color}
\usepackage{amsmath}
\usepackage{amssymb}
\usepackage{comment}
\usepackage{xcolor}

\usepackage[normalem]{ulem}

\DeclareRobustCommand{\erase}{\bgroup\markoverwith{\textcolor{red}{\rule[.5ex]{2pt}{0.4pt}}}\ULon}
 
\begin{document}


\title{Evaluating a Multi-Color Entangled-Photon Source for a Bosonic Silicon Quantum Circuit}

\author{Koki Nagamachi$^{1}$}

\author{Hiroki Yamashita$^{1}$}

\author{Mikio Fujiwara$^{2}$}

\author{Shigehito Miki$^{3}$}

\author{Hirotaka Terai$^{3}$}

\author{Takafumi Ono$^{1}$}
 \email{ono.takafumi@kagawa-u.ac.jp}

\affiliation{%
~\\
$^{1}$Program in Advanced Materials Science
Faculty of Engineering and Design,
Kagawa University,
2217-20 Hayashi-cho, Takamatsu, Kagawa
761-0396, Japan
}

\affiliation{%
$^{2}$Advanced ICT Research Institute, National Institute of Information and Communications Technology (NICT), Koganei, Tokyo 184-8795, Japan
}

\affiliation{%
$^{3}$Advanced ICT Research Institute, National Institute of Information and Communications Technology, 588-2 Iwaoka, Nishi, Kobe 651-2492, Japan
}

\date{\today}

\begin{abstract}
We evaluated a multi-color two-photon entangled state generated in silicon via spontaneous four-wave mixing (SFWM) as a potential source for bosonic integrated circuits. Spatially entangled photon states were created using a pair of silicon waveguides that produced signal and idler photons through SFWM, allowing us to observe quantum interference between them. Assuming that the frequencies of the multi-color photons were nearly identical, we characterized the generated quantum state by performing quantum state tomography on the bosonic system using a linear optical circuit. This study demonstrates the feasibility of using photon-pair sources generated in silicon via SFWM in bosonic optical circuits and highlights their potential for a wide range of applications in silicon-based optical quantum technologies.

\end{abstract}

\maketitle
\section{Introduction}
Silicon-based optical integrated circuits are promising platforms for large-scale quantum information processing \cite{Silverstone2016,rudolph2017optimistic,bunandar2018metropolitan,wang2020integrated,Moody_2022}. Their advantages include excellent scalability and controllability \cite{Zhang2022,Streshinsky:13}, as well as low optical loss in the telecommunications wavelength band \cite{MacFarlane2019,Hong:22,Bose2024}. Optical interferometers fabricated on silicon substrates offer compact and stable architectures, allowing for the integration of multiple stages to construct programmable linear optical circuits \cite{591665,Liu2004,soref2006past,Martijn2014,thomson2016roadmap,Atabaki2018,Near2021,Shekhar2024}. Importantly, photon pairs generated via spontaneous four-wave mixing (SFWM) in silicon serve as a valuable quantum resource, as both the photon source and the circuit can be monolithically integrated on a single silicon chip \cite{Clemmen:09,5325708,Xiong:11,Engin:13,Jiang:15,Hemsley2016,PhysRevLett.133.083803}.
Although the distinct frequencies of the signal and idler photons produced by SFWM typically inhibit quantum interference, it is still possible to observe interference by creating spatial or temporal superpositions of photon pairs from multiple SFWM sources \cite{silverstone2014chip,Wakabayashi:15,Ono:19}. This approach has enabled various experimental demonstrations, including time-bin entanglement \cite{Fujiwara:17,Samara:19,Ono2024}, the generation of large-scale quantum states \cite{wang2018multidimensional,Zhang2019,Lu2020}
, and the realization of a two-qubit quantum processor \cite{Silverstone2015,Harris2016,qiang2018large,Bao2023}.

In typical experiments where photon pairs are generated via SFWM in silicon, the frequency separation between the signal and idler photons is much smaller than their absolute optical frequencies \cite{ PhysRevApplied.18.034007, Sagawa2025}. Under these conditions, linear optical components such as multi-mode interference (MMI) couplers and phase shifters fabricated in silicon waveguides behave almost identically for both the signal and idler photons. As a result, their frequencies can be effectively regarded as nearly identical. This frequency indistinguishability constrains the dimensionality in which linear optical circuits can manipulate the quantum states generated by SFWM, limiting it to a subspace of the maximum possible Hilbert space.
Therefore, it is essential to understand the degree of controllability that a linear optical circuit can exert over a given quantum state. Such understanding not only guides the reduction of required optical components for practical quantum information processing but also facilitates accurate evaluation of the generated quantum states.

In this manuscript, we revisit the Hilbert space associated with quantum states generated via SFWM and manipulated through linear optics, exploring the potential of SFWM-generated photons as a source for bosonic quantum circuits. Specifically, we show that quantum states produced in a linear optical circuit with two spatial modes can be mapped onto a bosonic system described by a two-mode, same-color, two-photon state, which is mathematically equivalent to a spin-1 system.
Experimentally, we generated entangled photon pairs in a silicon waveguide using an interferometric SFWM source and injected them into a Mach–Zehnder interferometer (MZI) integrated on the same chip. By varying the phase of the interferometer, we observed high-visibility interference fringes in the two-photon output, confirming the coherent superposition of spatial-mode NOON-like states.
To quantitatively characterize the generated quantum states, we performed quantum state tomography within the spin-1 framework of the two-mode, two-photon Hilbert space. The quantum state was reconstructed by measuring the expectation values of spin-1 operators along five independent directions, enabling the evaluation of the state's purity, coherence, and internal consistency.
These results demonstrate that silicon photonic integrated circuits, combined with spin-based analysis and on-chip linear optics, provide a practical and scalable platform for implementing and characterizing bosonic quantum circuits.

\section{Quantum state generated via SFWM in a linear optical circuit}
\begin{figure}[b]
 \centering
 \includegraphics[width=\columnwidth]{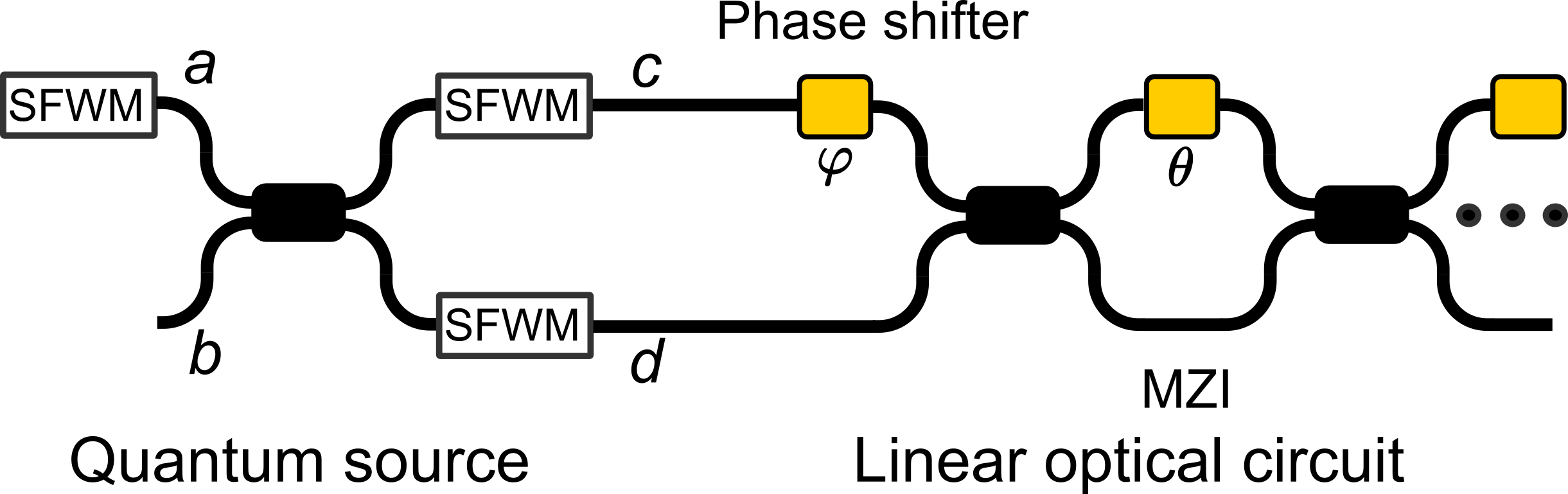}
 \caption{A conceptual diagram of a photonic quantum circuit combining spontaneous four-wave mixing and a linear optical circuit. When pump light is injected into a silicon waveguide, photon pairs are generated via spontaneous four-wave mixing. The generated quantum states are manipulated using a linear optical circuit that consists of an alternating sequence of MZI—each comprising two beam splitters and a phase shifter—and standalone phase shifters.
}
 \label{diagram}
\end{figure}

Figure \ref{diagram} illustrates a two-mode linear optical circuit that includes an interferometer with a variable phase shifter. The input state consists of two photons—a signal photon and an idler photon—generated via non-degenerate spontaneous four-wave mixing in silicon. As a result, the system under investigation is a two-mode, two-color, two-photon state. One possible basis for describing this system is the following set of four states:
\begin{equation}
\label{normal_basis}
{ |is;0 \rangle_{cd}, |i;s \rangle_{cd}, |s;i \rangle_{cd}, |0;is \rangle_{cd} }.
\end{equation}
Here, $s$ and $i$ denote signal and idler photons, respectively, while the subscripts $c$ and $d$ refer to spatial modes. In general, signal and idler photons generated by non-degenerate SFWM have different frequencies, and linear optical elements respond differently to each. However, when the frequency difference between the signal and idler photons is sufficiently small compared to their absolute frequencies, it is reasonable to assume that MMI couplers and phase shifters in optical circuits behave similarly for both photons. This approximation has been applied in various systems implemented with integrated linear optical circuits\cite{Silverstone2015,Sagawa2025}.

Specifically, the unitary transformations corresponding to the phase shifter $\hat{U}(\varphi)$ and the beamsplitter $\hat{U}_{\mathrm{BS}}$ in an optical circuit with spatial modes $c$ and $d$ can be expressed as follows:
\begin{eqnarray}
\label{unitary}
\hat{U}(\varphi) &=& {\rm exp} \left( -i \hat{c}^{\dagger} \hat{c} \varphi \right), \nonumber\\
\hat{U}_{\mathrm{BS}} &=& {\rm exp} \left( -i \pi \left( \hat{c}^{\dagger} \hat{d} + \hat{c} \hat{d}^{\dagger} \right)/2 \right),
\end{eqnarray}
where the operators $\hat{c}$ and $\hat{c}^{\dagger}$ represent the annihilation and creation operators for signal and idler photons in mode $c$, and $\hat{d}$ and $\hat{d}^{\dagger}$ denote the corresponding operators in mode $d$. It is important to note that the Hilbert space accessible via the unitaries in Eq.(\ref{unitary}) is a subset of the full space spanned by the basis in Eq.(\ref{normal_basis}), as these operations act identically on both frequencies.

As an example, by applying $\hat{U}_{\mathrm{BS}}$ to the states given in Eq.~(\ref{normal_basis}), one can derive the following set of three states as a possible basis for the system:
\begin{eqnarray}
|2;0\rangle_{cd} &:=& |is;0\rangle_{cd}, \nonumber\\
|1;1\rangle_{cd} &:=& \frac{1}{\sqrt{2}} \left( |i;s \rangle_{cd} + |s;i\rangle_{cd} \right), \nonumber\\
|0;2\rangle_{cd} &:=& |0;is\rangle_{cd},
\end{eqnarray}
where $|2;0\rangle_{cd}$ represents a state in which both photons are in mode $c$, $|1;1\rangle_{cd}$ corresponds to a state with one photon in each of the modes $c$ and $d$, and $|0;2\rangle_{cd}$ describes a state where both photons are in mode $d$. These three basis states correspond to those used in a two-mode, two-photon bosonic system, where the photons are indistinguishable. Any superposition of these three states that can be formed through a linear optical circuit is equivalent to a two-mode, two-photon bosonic state.

Under low-power conditions, photon pairs generated via SFWM are always found in states where both photons occupy the same spatial mode (e.g., $|is;0\rangle_{cd}$, $|0;is\rangle_{cd}$, or a superposition thereof). Therefore, when such states are injected into a linear optical circuit that does not contain frequency-dependent components, they can be effectively treated as two-mode, two-photon bosonic states.

In the following section, we experimentally investigate the quantum state generated via SFWM in a linear optical circuit. To facilitate the characterization of the resulting two-mode, two-photon quantum states, it is useful to introduce a set of operators that describe the symmetry and statistical structure of such states. These operators are defined as:
\begin{eqnarray}
\label{spin_operator}
\hat{L}_x &=& \frac{1}{2} \left( \hat{c}^{\dagger} \hat{d} + \hat{c} \hat{d}^{\dagger} \right), \nonumber \\
\hat{L}_y &=& -\frac{i}{2} \left( \hat{c}^{\dagger} \hat{d} - \hat{c} \hat{d}^{\dagger} \right), \nonumber \\
\hat{L}_z &=& \frac{1}{2} \left( \hat{c}^{\dagger} \hat{c} - \hat{d} \hat{d}^{\dagger} \right),
\end{eqnarray}
where $\hat{c}$ and $\hat{d}$ are annihilation operators corresponding to spatial modes $c$ and $d$, respectively. In a two-mode, one-photon system, these operators are mathematically equivalent to the Pauli spin operators of a spin-1/2 system. The states $|1;0\rangle_{cd}$ and $|0;1\rangle_{cd}$ act as spin-up and spin-down eigenstates of $\hat{L}_z$ with eigenvalues $+1/2$ and $-1/2$, respectively.

In the case of a two-mode, two-photon system, the analogy extends to a spin-1 system. The basis states ${ |2;0\rangle_{cd}, |1;1\rangle_{cd}, |0;2\rangle_{cd} }$, introduced in Eq.~(\ref{normal_basis}), are eigenstates of $\hat{L}_z$ with eigenvalues $+1$, $0$, and $-1$, respectively. These three states form a complete basis for a spin-1 representation of the SU(2) algebra in the bosonic Hilbert space defined by the two spatial modes. This spin analogy provides a framework for analyzing and visualizing quantum states in our system. In particular, expectation values of the operators $(\hat{L}_x, \hat{L}_y, \hat{L}_z)$ correspond to projections of the quantum state along the three spin axes, and thus contain complete information about the state when it resides in the symmetric two-photon subspace.

Experimentally, these spin projections can be accessed by applying appropriate unitary transformations to the input state—specifically, the phase shifter $\hat{U}$ and beamsplitter $\hat{U}_{\mathrm{BS}}$, as defined in Eq.(\ref{unitary}) and shown schematically in Fig.\ref{diagram}. By tuning the interferometer settings, one can rotate the measurement basis and extract the spin components, thereby enabling full quantum state tomography in the spin-1 (bosonic) subspace.

\section{Experimental setup}
\begin{figure}[t]
 \centering
 \includegraphics[width=\columnwidth]{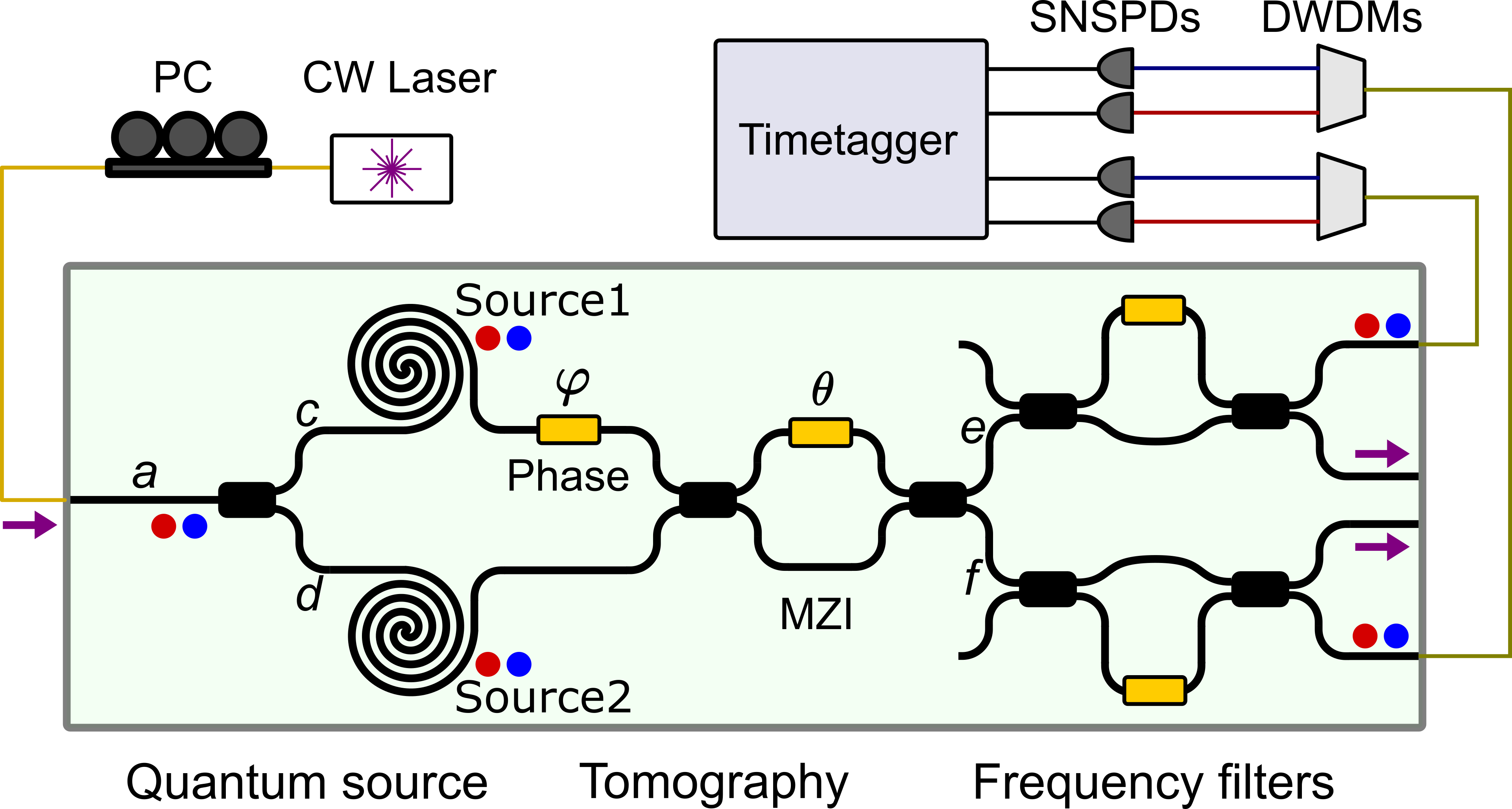}
 \caption{A schematic of the experimental setup. The laser source operating in the telecom wavelength band is first passed through a polarization controller (PC) to adjust the polarization, and then injected into the silicon waveguide. Photon pairs are generated via spontaneous four-wave mixing in modes $a$, $c$, and $d$, as indicated in the figure. Frequency filters were implemented using asymmetric interferometers.
}
 \label{experiment}
\end{figure}

In this framework, we used a silicon photonic integrated circuit to generate an entangled state in a two-mode, two-photon system and evaluated the resulting quantum state using the same circuit. The experimental setup is illustrated in Fig.\ref{experiment}. The integrated circuits were fabricated through a commercially available multi-project wafer (MPW) service. The chip dimensions are approximately 1 mm~$\times$ 2 mm, and the temperature was actively stabilized at 25°C using a Peltier element.

The experimental setup comprises two main components: a quantum state generation stage for producing spatially entangled photon pairs, and a measurement stage for performing quantum state tomography. As the excitation light source, we used a continuous-wave (CW) laser (CoBrite DX1, ID Photonics) operating at a wavelength of 1549.32 nm. The pump laser was coupled into the on-chip optical waveguide via a spot-size converter, with a coupling loss of approximately 4 dB. The estimated optical power coupled into the waveguide was approximately 5~mW.

We generated spatially entangled photon pairs using an interferometric setup consisting of two spiral waveguide sources\cite{silverstone2014chip}. The incident pump beam was first split into two spatial modes by a 50:50 beam splitter—implemented as a Y-branch 3-dB coupler—denoted by $\hat{U}_{\mathrm{BS}}$, and then guided into two spiral waveguides, one located in each arm of the interferometer. Each spiral waveguide used in the experiment has an estimated length of approximately 1.4~cm. Signal and idler photons were generated in both waveguides via SFWM.

Because the pump power was relatively low (on the order of a few milliwatts), the probability of generating photon pairs in both waveguides simultaneously was negligible. Thus, signal-idler photon pairs were generated in either mode $c$ or mode $d$, but not both at the same time. When coherent pump beams are used to drive the SFWM process, the resulting post-selected quantum state becomes a coherent superposition of two-photon states—one generated in mode $c$, $|2;0\rangle_{cd}$, and the other in mode $d$, $|0;2\rangle_{cd}$—as described by:
\begin{equation}
\label{ideal_NOON}
|\psi_{\mathrm{NOON}}\rangle = \frac{1}{\sqrt{2}} \left( |2;0\rangle_{cd} + |0;2\rangle_{cd} \right).
\end{equation}
This state is a two-photon NOON state, characterized by maximal path entanglement, and is commonly employed in quantum-enhanced phase estimation \cite{Nagata2007,Dowling2008,Ono2013}. 

As shown later in the quantum source section of Fig.2, the waveguide corresponding to mode $a$ has a length of approximately 1.5 mm. Photon pairs can also be generated via SFWM within this short waveguide segment. As a result, the actual quantum state injected into the tomography system is a superposition of the photon pair state generated in mode $a$ and the quantum state given in Eq.~(\ref{ideal_NOON}), and can be expressed as:
\begin{equation}
\label{generated_state}
|\psi_{in} \rangle = C(R) \left( \sqrt{R} \hat{U}_{\mathrm{BS}} |2;0\rangle_{ab} + \sqrt{1-R}|\psi_{\mathrm{NOON}} \rangle_{cd} \right),
\end{equation}
where $R$ denotes the probability that photon pairs are generated in mode $a$, and $C(R)$ is the normalization constant given by $1/\sqrt{1+2R(1-R)}$.

The generated quantum states are characterized via quantum state tomography using an on-chip analyzer composed of a phase shifter and an MZI. In the final stage of the analyzer, an on-chip frequency filter is implemented to remove the pump light while allowing the signal and idler photons to be analyzed as a bosonic quantum light source.

The free spectral range (FSR) of the on-chip filter is approximately 6.2 nm and was specifically designed to closely match the 200 GHz channel spacing of a C-band dense wavelength division multiplexing (DWDM) filter placed outside the photonic chip. In the experiment, signal and idler photons were observed at wavelengths approximately $\pm$ 9.3 nm from the pump laser wavelength of 1549.3 nm, corresponding to 1540.0 nm and 1558.3 nm, respectively. It is noteworthy that the wavelength shift of the signal and idler photons relative to the pump laser is less than 1\% of the central wavelength (1549~nm). The generated signal and idler photons were separated using a C-band DWDM filter and subsequently detected in coincidence using superconducting single-photon detectors. The quantum efficiency of the detectors used was approximately 60–80 \%, and the typical dark count rate was around 400 counts per second.

\begin{figure}[t]
 \centering
 \includegraphics[width=\columnwidth]{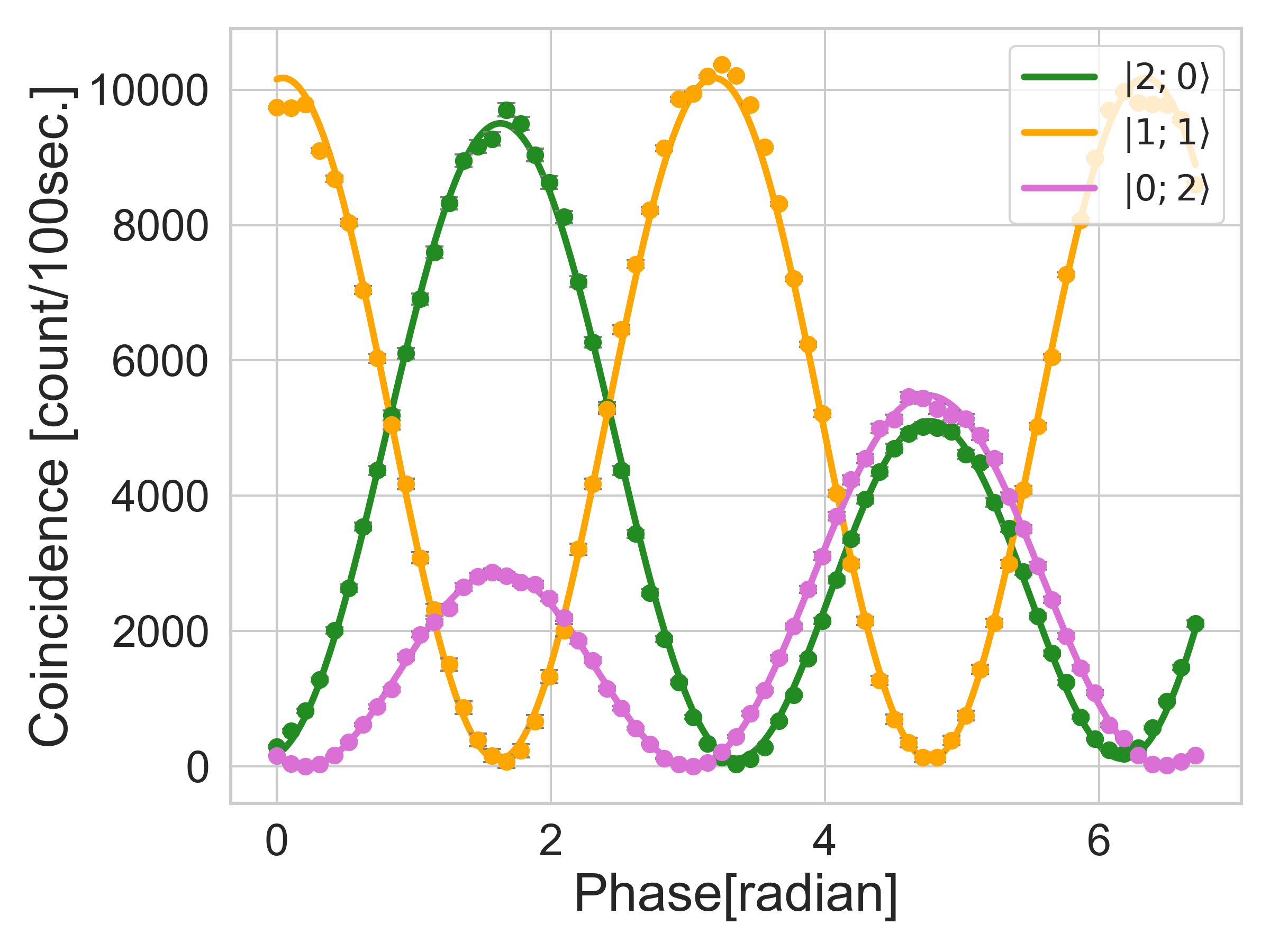}
 \caption{Interference fringes obtained using a silicon photonic integrated circuit. Although the error bars are too small to be visible in the figure, they were plotted assuming that the count statistics follow a Poisson distribution. The fitting was performed using expressions such as $N_{20} \times \langle 2;0| \hat{\rho}_{\mathrm{exp.}} |2;0\rangle$ for the measurement outcomes in the $|2;0\rangle$ basis, for example. Here, $N_{20}$ denotes a normalization constant, i.e., the total number of photons. The photon numbers obtained from the fitting were $N_{20} = 14830$, $N_{11} = 10535$, and $N_{02} = 8353$, which are used to calculate the probability distributions presented in the tomography results shown later.
}
 \label{fringe}
\end{figure}

To verify that the quantum state described in Eq.(\ref{generated_state}) was successfully generated, we conducted a quantum interference experiment using the output state. Specifically, we set the phase $\theta$ of the MZI to $\pi/2$, so that it acted as a 50:50 beam splitter. We then measured the coincidence counts between the signal and idler photons at the output while varying the phase $\varphi$ in mode $c$. The resulting interference fringes are shown in Fig.\ref{fringe}. The graphs display coincidence counts measured in the three bases ${ |2;0\rangle_{ef}, |1;1\rangle_{ef}, |0;2\rangle_{ef} }$. For the $|1;1\rangle_{ef}$ basis, we summed the measured counts from the $|i;s\rangle_{ef}$ and $|s;i\rangle_{ef}$ detection outcomes.

To quantitatively evaluate the interference visibility, we modeled the measured density matrix as a convex mixture of the ideal input state and a completely mixed state: $\hat{\rho}_{exp.} = V |\psi_{in} \rangle \langle \psi_{in} | + (1 - V) \hat{\rho}_{\mathrm{mix}}$, where $V$ denotes the visibility. By fitting the theoretical probabilities to the experimental data, we extracted both the visibility $V$ and the parameter $R$, which quantifies the contribution of photon pairs generated in mode $a$. From the best-fit curve, the observed visibilities in the three bases were $97.0 \pm 0.2\%$, $98.0 \pm 0.2\%$, and $99.8 \pm 0.2\%$, respectively. We also estimated $R \approx 0.024$, indicating that approximately 2.4\% of the photon pairs were generated in mode $a$.

\section{Quantum State Tomography of a Two-Mode Two-Photon State}
Finally, we performed quantum state tomography of the two-mode, two-photon system using our integrated circuit. As discussed in Section II, this system is equivalent to a spin-1 system. The quantum state of a spin-1 system is represented by a $3 \times 3$ density matrix, which is characterized by nine real parameters, including the normalization condition. Specifically, the density matrix can be expressed as \cite{PhysRevA.69.042108}:
\begin{equation}
\hat{\rho} = \frac{1}{3} \hat{I} + \frac{1}{2} \sum_{i=1}^3 \langle \hat{\lambda}_{1,i} \rangle \hat{\lambda}_{1,i} + \frac{1}{2} \sum_{i=1}^5 \langle \hat{\lambda}_{2,i} \rangle \hat{\lambda}_{2,i},
\end{equation}
where $\hat{\lambda}_{1,i}$ and $\hat{\lambda}_{2,i}$ are basis operators whose expectation values correspond to the first- and second-order moments of the optical spin operators ${ \hat{L}_i }$, which are linear combinations of the operators defined in Eq.(\ref{spin_operator}).

As shown in Ref.~\cite{PhysRevA.69.042108}, the density matrix can be fully reconstructed from the spin statistics up to second order by measuring along just five different spin directions. Following this theoretical proposal, we performed quantum state tomography using the following five spin directions:
\begin{align}
\hat{L}_1 = \hat{L}_x, ~~\hat{L}_2 = \hat{L}_y, ~~\hat{L}_3 = \frac{1}{\sqrt{2}} \left( \hat{L}_x + \hat{L}_y \right)\nonumber\\
 \hat{L}_4 = \frac{1}{\sqrt{2}} \left( \hat{L}_y + \hat{L}_z \right), ~~\hat{L}_5 = \frac{1}{\sqrt{2}} \left( \hat{L}_z + \hat{L}_x \right) ~
\end{align}
These five measurements were implemented by appropriately setting the phases $\theta$ and $\varphi$ in the interferometric circuit, as illustrated in Fig.~2.

\begin{table}[t]
\caption{Results of the tomography measurements. The plot shows the measurement probabilities corresponding to the five measurement directions defined in Eq. (4). Each probability distribution was also normalized so that the sum of probabilities for each measurement basis is equal to 1.
}
\label{tab:tomography_results}
\centering
\large
\setlength{\tabcolsep}{6pt}
\begin{tabular}{c|ccccc}
\hline
   & $L_1$ & $L_2$ & $L_3$ & $L_4$ & $L_5$ \\ \hline \hline
$P(20)$ & 0.333 & 0.017 & 0.189 & 0.306 & 0.394 \\ \hline
$P(11)$ & 0.027 & 0.926 & 0.366 & 0.418 & 0.034 \\ \hline
$P(02)$ & 0.655 & 0.019 & 0.445 & 0.254 & 0.572 \\ \hline
\end{tabular}
\normalsize
\end{table}

Table~I summarizes the output probabilities measured along the five spin directions. Here, $P(20)$, $P(11)$, and $P(02)$ denote the probabilities of observing the output states $|2;0\rangle$, $|1;1\rangle$, and $|0;2\rangle$, respectively. Each probability was obtained by dividing the number of counts observed in each measurement direction by the total number of photons estimated from the interference fringes fitted in Fig. 3. Based on these measurement results, we reconstructed the density matrix $\hat{\rho}_{\text{raw}}$ as follows:
\begin{align}
&\hat{\rho}_{\text{raw}} = 
\nonumber\\
&\begin{pmatrix}
0.575 & -0.122 - 0.039i & 0.466 - 0.143i \\
-0.122 + 0.039i & 0.010 & -0.106 + 0.037i \\
0.466 + 0.143i & -0.106 - 0.037i & 0.414
\end{pmatrix}
\end{align}
A graphical representation of the reconstructed density matrix is shown in Fig.4 (a). The eigenvalues of $\hat{\rho}_{\text{raw}}$ were calculated to be ${1.017, 0.016, -0.033}$. The appearance of a small negative eigenvalue is attributed to statistical noise in the experimental data, as the density matrix was reconstructed directly from raw measurements using Eq.(7), without applying maximum likelihood estimation (MLE). Despite a deviation of approximately 3\%, the result indicates that the reconstructed state is highly pure.

For reference, the density matrix obtained using MLE is given below:
\begin{align}
&\hat{\rho}_{\text{ML}} = 
\nonumber\\
&\begin{pmatrix}
0.532 & -0.100 - 0.029i & 0.465 - 0.108i \\
-0.100 + 0.029i & 0.032 & -0.090 + 0.049i \\
0.465 + 0.108i & -0.090 - 0.049i & 0.436
\end{pmatrix}
\end{align}
A graphical representation of the reconstructed density matrix is shown in Fig.4 (b). The trace distance between $\hat{\rho}_{\text{raw}}$ and $\hat{\rho}_{\text{ML}}$ was calculated to be $\mathrm{Tr}|\hat{\rho}_{\text{raw}} - \hat{\rho}_{\text{ML}}| \approx 0.06$. This small trace distance confirms the accuracy of the quantum state tomography and supports the reliability of the reconstruction method employed in this experiment.

\begin{figure}[t]
 \centering
 \includegraphics[width=\columnwidth]{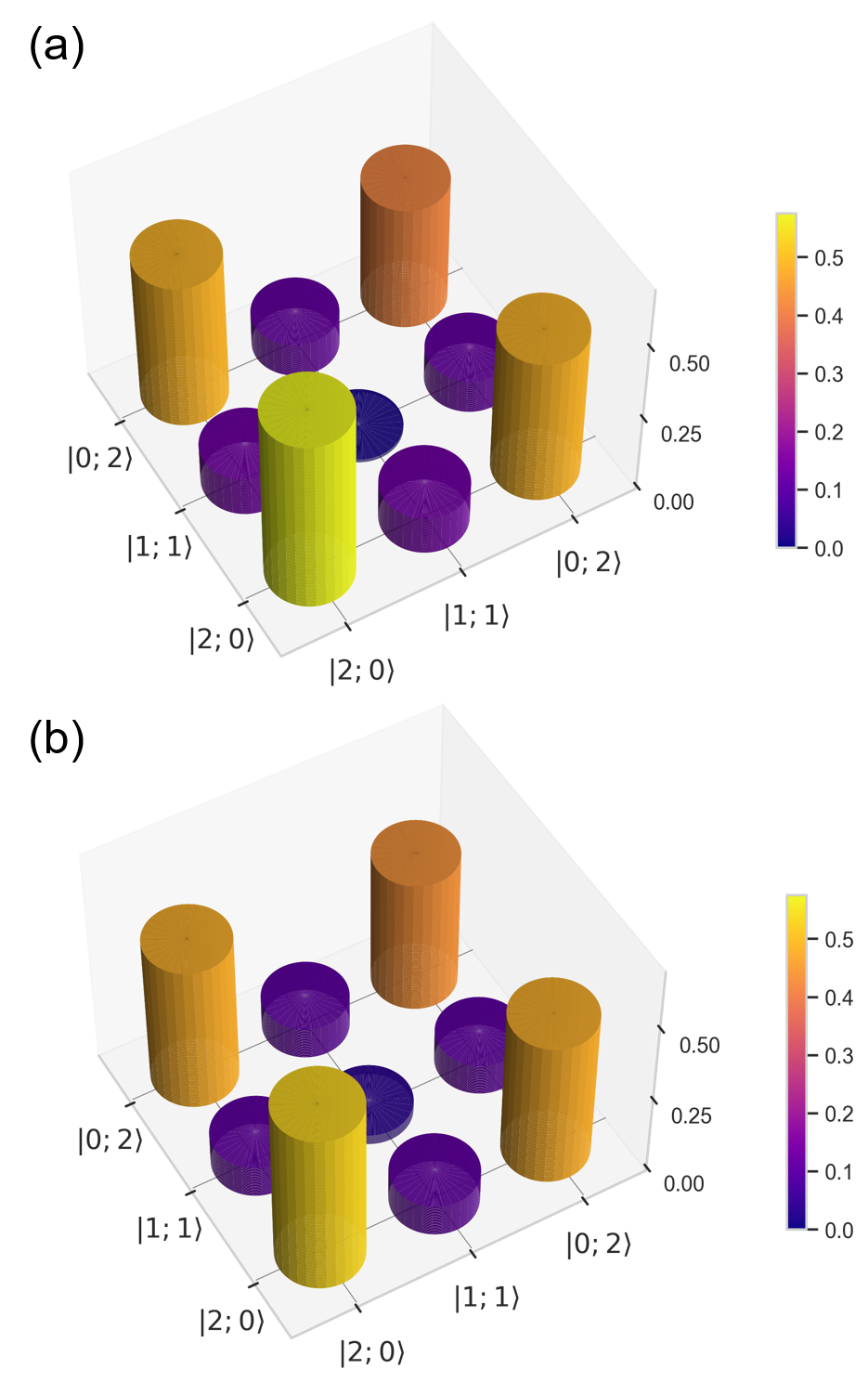}
 \caption{A graphical representation of the density matrix obtained from quantum state tomography. (a) Density matrix reconstructed via linear inversion, without applying maximum likelihood estimation. (b) Density matrix reconstructed using the maximum likelihood estimation method. While the density matrix elements are complex numbers, their absolute values are presented here.
}
 \label{tomography}
\end{figure}

Finally, we evaluate the experimental accuracy of our quantum state tomography based on the results presented in Table I. Since the five measurement operators defined in Eq.(8) satisfy specific linear relationships, the outcome of one measurement can be predicted from the others. These relations allow us to assess the internal consistency of the measured data, and thus the accuracy of the tomography~\cite{PhysRevA.69.042108}.

For example, the operator $\hat{L}_1$ can be expressed as a linear combination of $\hat{L}_3$, $\hat{L}_4$, and $\hat{L}_5$ as follows:
\begin{equation}
\hat{L}_1 = \frac{1}{\sqrt{2}} (\hat{L}_3 - \hat{L}_4 + \hat{L}_5).
\end{equation}
Using the experimental values in Table~I, the difference between the left- and right-hand sides is 0.012, corresponding to an error of approximately 1\%.
Similarly, $\hat{L}_2$ can be written as:
\begin{equation}
\hat{L}_2 = \frac{1}{\sqrt{2}} (\hat{L}_3 + \hat{L}_4 - \hat{L}_5),
\end{equation}
for which the experimental difference is 0.024, yielding an error of about 2\%.
In addition, $\hat{L}_z$ was independently measured and is related to the other operators by:
\begin{equation}
\hat{L}_z = \frac{1}{\sqrt{2}} (-\hat{L}_3 + \hat{L}_4 + \hat{L}_5).
\end{equation}
The experimentally obtained value of $\langle \hat{L}_z \rangle$ was 0.054, while the right-hand side evaluates to 0.081, resulting in a difference of 0.027, or approximately 3\% error. These results demonstrate that the linear relations among the measurement operators are well satisfied by the experimental data, indicating that our quantum state tomography was implemented with high accuracy.

\section{Conclusions}
In this work, we have demonstrated the generation and characterization of a two-mode, two-photon quantum state produced via SFWM in a silicon photonic integrated circuit. By exploiting the near-degeneracy of signal and idler photon frequencies, we treated the generated state as a bosonic two-photon system and manipulated it using on-chip linear optics. High-visibility quantum interference fringes were observed, and the state was quantitatively evaluated through spin-1-based quantum state tomography. The experimentally reconstructed density matrix showed high purity and internal consistency, with only a few percent deviation in the verification of spin operator relations. These results validate the use of spin-based tomography for bosonic states and demonstrate the precision and stability achievable in silicon photonic platforms. Our findings confirm that photon-pair sources based on SFWM in silicon, combined with on-chip linear optical circuits, provide a scalable and accurate platform for implementing bosonic quantum circuits. This architecture holds promise for advancing integrated quantum photonic technologies, particularly for applications requiring high-dimensional state control and compact circuit integration.

\textbf{\emph{Acknowledgements}}
This work was supported by JST PRESTO Grant No. JPMJPR1864, and JST ERATO Grant No. JPMJER2402, JSPS KAKENHI Grant Number JP24K00559, the Murata Science Foundation, the
 Shimazu Science Foundation and the Casio Science Promotion Foundation.

\bibliography{reference}

\end{document}